\documentclass[12pt]{article}
\begin{document}
\global\parskip 6pt
\newcommand{\be}{\begin{equation}}
\newcommand{\ee}{\end{equation}}
\newcommand{\bea}{\begin{eqnarray}}
\newcommand{\eea}{\end{eqnarray}}
\newcommand{\non}{\nonumber}

\begin{titlepage}
\hfill{hep-th/0208026} \vspace*{1cm}
\begin{center}
{\Large\bf Toroidal Black Holes and T-Duality}\\
\vspace*{2cm} Massimiliano Rinaldi\footnote{E-mail: massimiliano.rinaldi@ucd.ie}\\
\vspace*{.5cm}
{\em Department of Mathematical Physics\\
University College Dublin\\
Belfield, Dublin 4, Ireland}\\
\vspace{2cm}

\begin{abstract}
\noindent We consider the toroidal black holes that arise as a
generalization of the $AdS_5\times S^5$ solution of type IIB
supergravity. The symmetries of the horizon space allow T-duality
transformations that can be exploited to generate new inequivalent
black hole solutions of both type IIB and type IIA supergravity,
with non-trivial dilaton, $B$-field, and $RR$ forms. We examine
the asymptotic structure and thermodynamical properties of these
solutions.

\end{abstract}

\vspace{1cm} August 2002
\end{center}
\end{titlepage}
\vspace{1.5 cm}

\section {Introduction}
\noindent

\noindent In its simplest form, target space duality claims that a
bosonic closed string moving on a circle of radius $R$ is
physically equivalent to one moving on a circle of radius
$2\alpha'\hbar/R$, where $\alpha'$ is the string tension; for a
review, see \cite{Giveon}. This symmetry was then extended to many
compact directions, including the internal degrees of freedom
typical of heterotic string theory by Narain et al.
\cite{Narain1,Narain2}. For time-dependent backgrounds
\cite{Veneziano1}-\cite{Meissner}, the analogue of T-duality is
 the scale factor duality that relates a physically expanding and
a physically contracting universe. More generally, if the
background fields of a low energy effective action in string
theory are independent of $D$ coordinates, then this action shows
an $O(D,D)$ symmetry. The symmetry group is replaced by
$SO(D-1,1)$ if one of the coordinates is time-like. Hence, given a
solution of the equations of motion, one may generate a new one by
means of an $O(D,D)$ transformation. However, only the action of
the subgroup $O(D)\otimes O(D)/O(D)$ of $O(D,D)$ generates
physically inequivalent solutions \cite{Sen1}-\cite{Gasperini}.
T-duality is a particular case of an $O(D)\otimes O(D)$
transformation and if the $D$ coordinates are compact, the old and
the new solutions represent the same conformal field theory
\cite{Rocek}.

The aim of the present work is to investigate the symmetries of
toroidal black holes. These belong to a wider class of black hole
solutions to Einstein's equation with a negative cosmological
constant, called topological black holes \cite{Lemos}-\cite{Dias}.
They generalize the ordinary asymptotically flat Schwarzschild
black hole in $D$ dimensions to black holes that are locally
asymptotically anti-de Sitter such that the topology of the
horizon may be elliptic, toroidal, or hyperbolic. The toroidal
case clearly shows translational invariance along the compact
coordinates of the $(D-2)$-dimensional horizon space. The metric
is then independent of these $(D-2)$ coordinates, and the
effective action must exhibit an $O(D-2)$ symmetry group.
Moreover, since the metric is static, the action is also invariant
under an $SO(D-2,1)$ group of transformations as well. We will
investigate the inequivalent solutions obtained by the action of
the relevant subgroups on the toroidal black hole metric. In
particular, we will focus on the $5$-dimensional toroidal black
hole that can replace the $AdS_5$ sector of the $AdS_5\times S^5$
solution of type IIB supergravity \cite{Schwarz}. Therefore, we
must also study the action of the group on the anti-self dual
$5$-form that implements the model.

By performing a few preliminary computations, following
\cite{Sen1}-\cite{Gasperini}, we find that the action of the
$SO(3,1)\otimes SO(3,1)/SO(3,1)$ group always leads to metrics
with naked singularities. Hence, we will focus on the T-duality
transformations only, that belong to the $O(3)\otimes O(3)$
invariance group. In general, T-duality is a map between type IIB
and type IIA solutions, and we will explicitly see this through
the transformations of the spectrum of $RR$-forms. In strict
analogy with the duality between the uncharged and charged black
strings \cite{Horowitz2}, we will first boost the toroidal black
hole and then apply T-dualities in order to generate an axion
charge that is dual to the momentum. Also, we will check that the
entropy of the black hole is invariant under T-duality
\cite{Horowitz3}. Finally, we will see that the local asymptotic
structure of the black hole changes from anti-de Sitter to warped
products of compactified Minkowski space-times and hyperbolic
spaces.

\section {Toroidal Black Holes}
Toroidal black holes belong to a larger class of topological black
holes that arise as solutions of the Einstein equations with a
negative cosmological constant \cite{Lemos}-\cite{Birmingham}, see
also \cite{Lemos2} and \cite{Dias} for the charged counterparts.
They are Einstein space-times which are locally asymptotically
anti-de Sitter. The horizon may have elliptic, toroidal, or
hyperbolic topology as opposed to asymptotically flat black holes,
for which only spherical topology is possible. The toroidal case
has the $D$-dimensional metric \be
ds^2=-f(r)dt^2+f(r)^{-1}dr^2+\frac{r^2}{l^2}\delta_{\alpha\beta}dx^{\alpha}dx^{\beta},
\ee where \be f(r)=\frac{r^2}{l^2}-\frac{2M}{r^{D-3}}. \ee Here,
$M$ is the mass parameter, $l$ is the $AdS$ radius, and
$x^{\alpha}$ ($\alpha=1,\ldots ,D-2$) are the compact coordinates
parameterizing the torus horizon. The horizon is located at
$r_+^{D-1}=2Ml^2$. For $D=5$, this metric may replace the $AdS_5$
sector of the $AdS_5\times S^5$ solution of type IIB supergravity
\cite{Schwarz}, see also \cite{Strominger,Duff}. Then, the
$10$-dimensional metric is \be\label{tormetric}
ds^2=-f(r)dt^2+f(r)^{-1}dr^2+\frac{r^2}{l^2}\left[dx_1^2+dx_2^2+dx_3^2\right]+l^2d\Omega_5^2
.\ee This solution must be supplemented with an anti-self dual
$5$-form
\begin{eqnarray}\label{5form}
F_{\mu_1...\mu_5} &=& +\frac{4}{l}\varepsilon_{\mu_1...\mu_5},  \nonumber \\
F_{m_1...m_5}&=& -\frac{4}{l}\varepsilon_{m_1...m_5},
\end{eqnarray} in order to satisfy the fundamental equation \cite{Schwarz} \be R_{MN}=\frac{1}{6\cdot4^2}
F_M^{\hspace{3mm}A_2A_3A_4A_5}F_{NA_2A_3A_4A_5}. \ee With these
conventions, $\varepsilon_{01\ldots 89}=-\sqrt{-\det G}$,
$\mu=t,r,x_1,x_2,x_3,$ and the $m$'s are the indices of the
coordinates on the $5$-sphere $S^5$. Upper case Latin letters
label all the coordinates. The Hodge operator * applied to a
$p$-form is defined such that ${}^{**}F^{(p)}=(-1)^{p+1}F^{(p)}$.

It is clear that the metric (\ref{tormetric}) is static and
translationally invariant along the space-like compact directions
$x_1,x_2$, and $x_3$. Thus, we can generate new inequivalent
solutions by means of an $O(3)\otimes O(3)/O(3)$ or an
$SO(3,1)\otimes SO(3,1)/SO(3,1)$ group of transformations. As
mentioned above, in the latter case we always obtain metrics with
naked singularities, and we will focus on T-duality
transformations that belong to the more general $O(3)\otimes O(3)$
group. Since $x_1,x_2$, and $x_3$ are compact, all the new
solutions that we obtain represent the same conformal theory
\cite{Rocek}.

\section{Application of T-duality}

The T-duality transformations for the Neveu-Schwarz sector of an
effective theory are well known. Given a metric $G_{MN}$, a
$B$-field $B_{MN}$, and dilaton $\phi$, independent of the
coordinate $x$, the transformation rules are
\cite{Buscher,Buscher2}
\begin{eqnarray}\label{tduality}\nonumber
\tilde G_{xx}&=&1/G_{xx}, \qquad \tilde G_{xN}=B_{xN}/G_{xx},
\\ \nonumber \tilde
G_{MN}&=&G_{MN}-(G_{xM}G_{xN}- B_{xM}B_{xN})/G_{xx},
\\ \nonumber \tilde B_{xM}&=&G_{xM}/G_{xx}, \qquad
\tilde B_{MN}=B_{MN}-2G_{x[M}B_{N]x}/G_{xx},
\\ 2\tilde \phi &=&2\phi-\ln G_{xx},
\end{eqnarray} where $M$ and $N$ run over all the coordinates except
$x$. Here, $T_{[MN]}=(1/2!)\left(T_{MN}-T_{NM}\right)$. The type
IIA/IIB T-duality transformation rules for the $RR$-form fields
were derived more recently in \cite{Bergshoeff1,Bergshoeff2}, and
put in a general form by Hassan in \cite{Hassan1} (see also
\cite{Hassan2} for the $SO(D,D)$ transformation rules). According
to \cite{Hassan1}, the components of an $RR$-form $F$ which is
independent of $x$ are related to the components of its T-dual
$\tilde F$ by
\begin{eqnarray}\label{tdualityform}\nonumber \tilde F^{(n)}_{x M_2\ldots
M_n}&=&F^{(n-1)}_{ M_2\ldots M_n}-(n-1)G^{xx}G_{x[
M_2}F^{(n-1)}_{x M_3\ldots M_n]},\\\nonumber\\ \tilde F^{(n)}_{
M_1\ldots M_n}&=&F^{(n+1)}_{x M_1\ldots M_n}-nB_{x[M_1}\tilde
F^{(n)}_{x M_2\ldots M_n]}. \end{eqnarray} In this formalism, the
index $M$ runs over all the coordinates except $x$, the
anti-symmetrization symbols on the right hand sides do not involve
the index $x$, and the upper index of the form indicates its rank.
Following closely the method explained in \cite{Horowitz2} (see
also \cite{Horowitz4,Horowitz}), we first perform a boost by the
change of coordinates $x_1\rightarrow x_1 \cosh \alpha+t \sinh
\alpha$, $t\rightarrow t\cosh \alpha+x_1\sinh \alpha$. Then, we
apply a T-duality on the background fields along $x_1$ using
(\ref{tduality}). The resulting metric, dilaton field, and
$B$-field take the form
\begin{eqnarray}\label{dual1}\nonumber
ds^2&=&-\frac{r^2}{l^2}h(r)g(r)^{-1}dt^2+\frac{l^2}{r^2}h(r)^{-1}dr^2+
\frac{l^2}{r^2}g(r)^{-1}dx_1^2+\frac{r^2}{l^2}\left[dx_2^2+dx_3^2\right]+l^2d\Omega_5^2,
\\\nonumber\\\nonumber
B_{tx_1}&=&-\frac{q}{r^4}g(r)^{-1}, \\\nonumber\\
e^{-2\phi}&=&\frac{r^2}{l^2}g(r),  \end{eqnarray} where
\begin{eqnarray} \nonumber h(r)&=&1-\frac{m}{r^4}+\frac{q^2}{mr^4},
\\\nonumber\\ g(r)&=&1+\frac{q^2}{mr^4}. \end{eqnarray}
The axion charge is given by $q=2Ml^2\cosh \alpha \sinh \alpha $,
and the rescaled mass is $m=2Ml^2\cosh^2 \alpha$. In analogy with
the charged black string found in \cite{Horowitz2}, the horizon is
unique and its location is unchanged: $r_+^4=(m^2-q^2)/m=2Ml^2$.
Also, in the extremal limit $M\rightarrow 0$, $\alpha\rightarrow
\infty$ (such that $m=q$) one can check that the horizon
disappears and the metric is no
 longer singular. However, as opposed to \cite{Horowitz2}, in the limit
$\alpha\rightarrow 0$ we do not recover the original metric,
because of the term $r^2/l^2$ that multiplies the horizon
sub-metric.

In the large $r$ limit, the metric of the black hole reduces to
\be ds^2\approx
\frac{r^2}{l^2}\left[-dt^2+dx^2_2+dx_3^2\right]+\frac{l^2}{r^2}\left[dr^2+dx_1^2\right]
\ee which looks singular as $r\rightarrow\infty$. However, it can
be checked that the scalar curvature of the black hole tends to
$G_{\mu\nu}R^{\mu\nu}=-8/l^2$ and all the scalar curvature
polynomials are finite (and constant) in the limit. This is not
surprising since the topology of the space-time for large $r$ is a
warped product of a $3$-dimensional Minkowski space-time,
compactified along $x_2$ and $x_3$, and a $2$-dimensional
hyperbolic space compactified along $x_1$.

In general, T-duality is a map between type IIA and type IIB
supergravity solutions. Hence, we expect that the anti-self dual
$5$-form is mapped to forms with even rank. This can be readily
verified by means of the transformation rules
(\ref{tdualityform}), and we find
\begin{eqnarray}\label{forms1}\nonumber
F^{(4)}_{\quad\mu_1\ldots\mu_4}&=&\frac{4r}{l^2}\sqrt{g(r)}\varepsilon_{\mu_1\ldots\mu_4},\qquad
\mu\neq x_1 \\\nonumber\\ F^{(6)}_{\quad A_1\ldots
A_6}&=&-\frac{4r}{l^2}\sqrt{g(r)}\varepsilon_{A_1\ldots
A_6},\qquad A_i\neq t,r,x_2,x_3, \end{eqnarray} where
$\varepsilon$ denotes the volume form of the corresponding
subspace. It can be verified that $F^{(4)}={}^{*}F^{(6)}$ and
$F^{(6)}=-{}^{*}F^{(4)}$. Thus, we have obtained a non-trivial
solution for type IIA supergravity with a black hole sector,
$B$-field, dilaton, and $RR$ forms.

The above solution is still translationally invariant along
$x_1,x_2$, and $x_3$. For example, we could apply another boost
and a T-duality along $x_2$ to generate an axion charge
independent of $q$. For simplicity, we restrict our attention to
T-duality only. First, a T-duality transformation along $x_2$
yields
\begin{eqnarray}\label{dual2}\nonumber ds^2&=&-\frac{r^2}{l^2}h(r)g(r)^{-1}dt^2+\frac{l^2}{r^2}h(r)^{-1}dr^2+
\frac{l^2}{r^2}\left[g(r)^{-1}dx_1^2+dx^2_2\right]+\frac{r^2}{l^2}dx_3^2+l^2d\Omega_5^2,
\\\nonumber\\\nonumber
B_{tx_1}&=&-\frac{q}{r^4}g(r)^{-1}, \\\nonumber\\
e^{-2\phi}&=&\frac{r^4}{l^4}g(r).  \end{eqnarray} By computing the
forms, we find a $3$-form and a $7$-form,
\begin{eqnarray}\label{forms2}\nonumber
F^{(3)}_{\quad\mu_1\mu_2\mu_3}&=&\frac{4r^2}{l^3}\sqrt{g(r)}\varepsilon_{\mu_1\mu_2\mu_3},\qquad
\mu\neq x_1,x_2, \\\nonumber\\ F^{(7)}_{\quad A_1\ldots
A_7}&=&-\frac{4r^2}{l^3}\sqrt{g(r)}\varepsilon_{A_1\ldots
A_7},\qquad A_i\neq t,r,x_3. \end{eqnarray} It can be verified
that $F^{(3)}={}^{*}F^{(7)}$, $F^{(7)}={}^{*}F^{(3)}$, and we see
that this is again a solution for type IIB supergravity. The
horizon location is unaltered and, in the extremal limit $m=q$,
the singularity disappears. In the large $r$ limit, the black hole
metric has the topology of a warped product of $2$-dimensional
Minkowski space-time, with $x_3$ compact, and a $3$-dimensional
hyperbolic space compactified along $x_1$ and $x_2$. Again, there
are no singularities as $r\rightarrow \infty$ and the scalar
curvature tends to a constant.

As a final step, we apply a T-duality along $x_3$ to the solution
above. The metric, the $B$-field, and the dilaton are given by
\begin{eqnarray}\label{dual3}\nonumber ds^2&=&-\frac{r^2}{l^2}h(r)g(r)^{-1}dt^2+\frac{l^2}{r^2}h(r)^{-1}dr^2+
\frac{l^2}{r^2}\left[g(r)^{-1}dx_1^2+dx^2_2+dx^2_3\right]+l^2d\Omega_5^2,
\\\nonumber\\\nonumber
B_{tx_1}&=&-\frac{q}{r^4}g(r)^{-1}, \\\nonumber\\
e^{-2\phi}&=&\frac{r^6}{l^6}g(r).  \end{eqnarray} The previous
forms are mapped into a $2$-form and an $8$-form
\begin{eqnarray}\label{forms3}\nonumber
F^{(2)}_{\quad\mu_1\mu_2}&=&\frac{4r^3}{l^4}\sqrt{g(r)}\varepsilon_{\mu_1\mu_2},\qquad
\mu\neq x_1,x_2,x_3, \\\nonumber\\ F^{(8)}_{\quad A_1\ldots
A_8}&=&-\frac{4r^3}{l^4}\sqrt{g(r)}\varepsilon_{A_1\ldots
A_8},\qquad A_i\neq t,r, \end{eqnarray} that satisfy the type IIA
duality condition $F^{(2)}={}^{*}F^{(8)}$ and
$F^{(8)}=-{}^{*}F^{(2)}$. All the previous characteristics are
unchanged, i.e. the horizon location and the the absence of
curvature singularities as $r\rightarrow \infty$. Also, the $r=0$
curvature singularity disappears in the extremal limit $m=q$. In
particular, we see that in this limit the black hole metric
reduces to a warped product of time and a $4$-dimensional
hyperbolic space compactified along $x_1,x_2$ and $x_3$, with a
scalar curvature $G_{\mu\nu}R^{\mu\nu}\rightarrow -8/l^2$.

According to the general formalism, these duality transformations
generate solutions of type IIA or type IIB supergravity. In
particular, we explicitly checked that the fields
(\ref{dual3})-(\ref{forms3}) are a non-trivial solution of the
type IIA supergravity. Indeed, these are solutions for the
equations of motion \cite{Polchinski}
\begin{eqnarray}\nonumber
0&=&R_{AB}+2\nabla_A\nabla_B\phi-\frac{1}{4}H_A^{\;\; CD}H_{BCD}+
\frac{1}{4\cdot2!}e^{2\phi}F_{CD}F^{CD}G_{AB}-\frac{1}{2}e^{2\phi}F_A^{\;\;
C}F_{BC},
\\\nonumber\\\nonumber 0&=&\nabla_M\left(e^{-2\phi}H^{M}_{\;\;PQ}\right),
\\\nonumber\\
0&=&4\nabla^2\phi-4\left(\nabla\phi\right)^2+R-\frac{1}{12}H_{ABC}H^{ABC},
\end{eqnarray} where $H_{MNP}=3!\:\partial_{[M}
B_{NP]}$. These equations are obtained by variation of the
effective type IIA action \be\label{action}
S=\frac{1}{2\kappa^2}\int
d^{10}x\sqrt{-G}\left\{e^{-2\phi}\left[R+4\nabla^2\phi-
\frac{1}{12}H_{MNP}H^{MNP}\right]-\frac{1}{4}F_{MN}F^{MN}\right\},
\ee with vanishing $4$-form.

In general, if an asymptotically flat low energy string theory
solution has a horizon and at least one space-like symmetry, then
the dual solution also has a horizon, with the same area (in
Einstein frame), and the same temperature \cite{Horowitz3}. The
horizon area is invariant in our case as well. After the boost of
the metric (\ref{tormetric}), the area is $A=(2\pi
r_+/l)^3\sqrt{g(r_+)}$. The $5$-dimensional black hole metric in
Einstein frame $ds^2_{(E)}$ is related to the one in string frame
$ds^2_{(ST)}$ through
$ds^2_{(E)}=exp\left(-4\phi/3\right)ds^2_{(ST)}$. In particular,
let $H_{ij}$ ($i=1,2,3$) be the horizon metric of (\ref{dual3}) in
Einstein frame. Then \be
H_{ij}dx^{i}dx^{j}=\frac{r^2}{l^2}g(r)^{-1/3}dx_1^2
+\frac{r^2}{l^2}g(r)^{2/3}\left(dx_2^2+dx_3^2\right). \ee The area
is then $(2\pi)^3\sqrt{\det H(r_+)}$ which coincides with $A$.
Hence, the black hole and its dual have the same entropy. It is
straightforward to check that this is true for the metrics
(\ref{dual1}) and (\ref{dual2}) as well.

\section{Conclusions}
The application of T-duality transformations to the toroidal black
hole generates new black hole solutions in the context of type IIA
and type IIB supergravity with non trivial dilaton, $B$-field, and
$RR$ forms. The analogy with the duality between the charged and
uncharged black string is strict, and all the new metrics have a
unique horizon with unchanged location and entropy. We also
analyzed the asymptotic structure and showed that it takes the
form of a warped product of compactified hyperbolic spaces and
Minkowski space-times. As opposed to the original toroidal black
hole, the new metrics are not Einstein space-times, and the Ricci
scalar is constant only in the large $r$ limit.

In the original solution (\ref{tormetric}) and (\ref{5form}), the
effect of the $5$-form is to produce an effective cosmological
constant. One might wonder whether a similar mechanism is valid
for the dualized solutions. Namely, one would like to check if,
for example, the solution (\ref{dual3}) can be generated by
replacing the $RR$-forms (\ref{forms3}) by a constant term in the
action (\ref{action}). Preliminary computations (with $q=0$),
indicate that this mechanism does not work. Further investigations
may be worthy. Finally, we note that topological black holes with
hyperbolic topology \cite{Birmingham} show translational
invariance as well and it would be interesting to apply T-duality
transformations, in analogy with the toroidal case.

\noindent {\bf \large Acknowledgements}. We would like to thank
Danny Birmingham for valuable discussions. This work was supported
by Enterprise Ireland grant BR/1999/031, and P.E. 2000/2002 from
Bologna University.


\begin{thebibliography}{99}



\bibitem {Giveon} A.~Giveon, M.~Porrati and E.~Rabinovici,
Phys.\ Rept.\  {\bf 244} (1994) 77; hep-th/9401139.

\bibitem{Narain1}
K.~S.~Narain, Phys.\ Lett.\ B {\bf 169}, 41 (1986).

\bibitem{Narain2}
K.~S.~Narain, M.~H.~Sarmadi and E.~Witten, Nucl.\ Phys.\ B {\bf
279}, 369 (1987).

\bibitem{Veneziano1}
G.~Veneziano, Phys.\ Lett.\ B {\bf 265} (1991) 287.

\bibitem{Veneziano2}
K.~A.~Meissner and G.~Veneziano, Phys.\ Lett.\ B {\bf 267} (1991)
33.

\bibitem{Meissner}
K.~A.~Meissner and G.~Veneziano, Mod.\ Phys.\ Lett.\ A {\bf 6}
(1991) 3397; hep-th/9110004.

\bibitem{Sen1}
A.~Sen, Phys.\ Lett.\ B {\bf 271} (1991) 295.

\bibitem{Sen2}
A.~Sen, Phys.\ Lett.\ B {\bf 274} (1992) 34; hep-th/9108011.

\bibitem{Sen3}
S.~F.~Hassan and A.~Sen, Nucl.\ Phys.\ B {\bf 375} (1992) 103;
hep-th/9109038.

\bibitem{Gasperini}
M.~Gasperini, J.~Maharana and G.~Veneziano, Phys.\ Lett.\ B {\bf
272} (1991) 277.

\bibitem{Rocek} M.~Ro\u cek and E.~Verlinde,
Nucl.\ Phys.\ B {\bf 373} (1992) 630; hep-th/9110053.

\bibitem{Lemos}
J.P. Lemos, Phys.\ Lett.\ B {\bf 353}, 46 (1995); gr-qc/9404041.

\bibitem{Huang}
C.~G.~Huang and C.~B.~Liang, Phys.\ Lett.\ A {\bf 201} (1995) 27.

\bibitem{Cai}
R.~G.~Cai and Y.~Z.~Zhang, Phys.\ Rev.\ D {\bf 54} (1996) 4891;
gr-qc/9609065.

\bibitem{Beng} S. {\AA}minneborg, I. Bengtsson, S. Holst,
and P. Peld\'{a}n, Class. Quantum Grav. 13 (1999) 2707;
gr-qc/9604005.

\bibitem{Mann} R.B. Mann, Class. Quantum Grav. 14 (1997) L109;
gr-qc/9607071.

\bibitem{Vanzo} L. Vanzo, Phys. Rev. D56 (1997) 6475; gr-qc/9705004.

\bibitem{Brill} D.R. Brill, J. Louko, and P. Peld\'{a}n,
Phys. Rev. D56 (1997) 3600; gr-qc/9705012.

\bibitem{Birmingham} D. Birmingham,
Class. Quantum Grav. 16 (1999) 1197; hep-th/9808032.

\bibitem{Lemos2}
J.~P.~Lemos and V.~T.~Zanchin, Phys.\ Rev.\ D {\bf 54} (1996) 3840
; hep-th/9511188.

\bibitem{Dias}
O.~J.~Dias and J.~P.~Lemos, Class.\ Quant.\ Grav.\  {\bf 19}
(2002) 2265; hep-th/0110202.


\bibitem{Schwarz}
J.~H.~Schwarz, Nucl.\ Phys.\ B {\bf 226} (1983) 269.

\bibitem {Horowitz2} J.~H.~Horne, G.~T.~Horowitz and A.~R.~Steif,
Phys.\ Rev.\ Lett.\  {\bf 68} (1992) 568; hep-th/9110065.

\bibitem{Horowitz3} G.~T.~Horowitz and D.~L.~Welch,
Phys.\ Rev.\ D {\bf 49} (1994) 590; hep-th/9308077.

\bibitem{Strominger}
G.~T.~Horowitz and A.~Strominger, Nucl.\ Phys.\ B {\bf 360} (1991)
197.

\bibitem{Duff}
M.~J.~Duff and J.~X.~Lu, Phys.\ Lett.\ B {\bf 273} (1991) 409.

\bibitem{Buscher} T.~H.~Buscher, Phys.\ Lett.\ B {\bf 201} (1988) 466.

\bibitem{Buscher2} T.~H.~Buscher, Phys.\ Lett.\ B {\bf 194} (1987) 59.

\bibitem{Bergshoeff1} E.~Bergshoeff, C.~M.~Hull and T.~Ortin,
Nucl.\ Phys.\ B {\bf 451} (1995) 547; hep-th/9504081.

\bibitem{Bergshoeff2} E.~Bergshoeff, M.~de Roo, M.~B.~Green, G.~Papadopoulos and P.~K.~Townsend,
Nucl.\ Phys.\ B {\bf 470} (1996) 113; hep-th/9601150.

\bibitem{Hassan1}S.~F.~Hassan,
Nucl.\ Phys.\ B {\bf 568} (2000) 145; hep-th/9907152.

\bibitem{Hassan2}
S.~F.~Hassan, Nucl.\ Phys.\ B {\bf 583} (2000) 431;
hep-th/9912236.

\bibitem{Horowitz4} J.~H.~Horne and G.~T.~Horowitz,
Nucl.\ Phys.\ B {\bf 368} (1992) 444; hep-th/9108001.

\bibitem{Horowitz} G.~T.~Horowitz, ``The dark side of string
theory: Black holes and black strings,''; hep-th/9210119.

\bibitem{Polchinski}
J.~Polchinski, ``String Theory Vol.I-II,'' Cambridge University
Press, Cambridge, 1998.


\end{thebibliography}
\end{document}